\documentclass[10pt,journal,journal]{IEEEtran}
\usepackage{pgfplots}

\usepackage{scalerel}
\usepackage{tikz}
\usetikzlibrary{svg.path}
\usepackage{setspace}
\usepackage{xcolor}

\usepackage{color}
\usepackage{colortbl}
\usepackage{multirow}

\definecolor{orcidlogocol}{HTML}{A6CE39}

\definecolor{darkgreen}{RGB}{0,80,0}

\definecolor{orange}{RGB}{255,127,0}

\tikzset{
    orcidlogo/.pic={
        \fill[orcidlogocol] svg{M256,128c0,70.7-57.3,128-128,128C57.3,256,0,198.7,0,128C0,57.3,57.3,0,128,0C198.7,0,256,57.3,256,128z};
        \fill[white] svg{M86.3,186.2H70.9V79.1h15.4v48.4V186.2z}
        svg{M108.9,79.1h41.6c39.6,0,57,28.3,57,53.6c0,27.5-21.5,53.6-56.8,53.6h-41.8V79.1z M124.3,172.4h24.5c34.9,0,42.9-26.5,42.9-39.7c0-21.5-13.7-39.7-43.7-39.7h-23.7V172.4z}
        svg{M88.7,56.8c0,5.5-4.5,10.1-10.1,10.1c-5.6,0-10.1-4.6-10.1-10.1c0-5.6,4.5-10.1,10.1-10.1C84.2,46.7,88.7,51.3,88.7,56.8z};
    }
}

\newcommand\orcidicon[1]{\href{https://orcid.org/#1}{\mbox{\scalerel*{
                \begin{tikzpicture}[yscale=-1,transform shape]
                \pic{orcidlogo};
                \end{tikzpicture}
            }{|}}}}

\usepackage{float}

 \usepackage{graphicx}
\usepackage[T1]{fontenc}
\usepackage{url}
\usepackage{mathtools}

 \usepackage{framed, color}
\definecolor{shadecolor}{rgb}{0.79,0.87,1}

\usepackage{amsmath}
\usepackage{amssymb}                        
\usepackage{amsfonts}
\usepackage{mathtools}
\usepackage{pstricks}
\usepackage{pst-all}
\usepackage{multirow}
\usepackage{slashbox}
\usepackage{url}
\usepackage{algorithmic}
\usepackage{enumitem}
\usepackage[ruled,vlined]{algorithm2e}
\usepackage{tikz}
\usetikzlibrary{shapes,arrows}
\usetikzlibrary{shapes, arrows, positioning, calc}
\usepackage{xcolor}

\usepackage{multicol, blindtext}
\usepackage{etoolbox}
\usepackage{pgfplots}
\usepackage{pgfplotstable}
\usepgfplotslibrary{groupplots}
\pgfplotsset{compat=1.7}
        
\SetKwInput{KwData}{Input}
\SetKwInput{KwResult}{Output}
\SetKwInput{KwSteps}{Steps}

\usepackage{pifont}

\def\thebibliography#1{\section*{References\markboth
 {REFERENCES}{REFERENCES}}\list
 {[\arabic{enumi}]}{\settowidth\labelwidth{[#1]}\leftmargin\labelwidth
 \advance\leftmargin\labelsep
 \usecounter{enumi}}
 \def\newblock{\hskip .11em plus .33em minus -.07em}
 \sloppy
 \sfcode`\.=1000\relax}

\usepackage[utf8]{inputenc}
\usepackage{amsthm}
\theoremstyle{plain}

\newtheorem{corollary}{Corollary}
\newtheorem{example}{Example}
\newtheorem{fact}{Fact}

\newtheorem{lemma}{Lemma}
\newtheorem{definition}{Definition}

\usepackage{hyperref}
\usepackage{listings}

\onecolumn

\begin{document}
\title{Compression Optimality of Asymmetric Numeral Systems}
\author{
Josef Pieprzyk$^{\textsuperscript{\orcidicon{0000-0002-1917-6466}}}$,
Jarek Duda$^{\textsuperscript{\orcidicon{0000-0001-9559-809X}}}$,
Marcin Paw\l{}owski$^{\textsuperscript{\orcidicon{0000-0002-5145-9220}}}$,
Seyit Camtepe$^{\textsuperscript{\orcidicon{0000-0001-6353-8359}}}$,
Arash Mahboubi$^{\textsuperscript{\orcidicon{0000-0002-0487-0615}}}$,
Pawe\l{} Morawiecki$^{\textsuperscript{\orcidicon{0000-0003-3349-8645}}}$
\IEEEcompsocitemizethanks{
\IEEEcompsocthanksitem
\hrule{}{}
\vspace{2mm}
Josef Pieprzyk is with
Institute of Computer Science, Polish Academy of Sciences, Warsaw, Poland and
Data61, CSIRO, Sydney, Australia, email: {\sl josef.pieprzyk@csiro.au}
\IEEEcompsocthanksitem
Jarek Duda is with 
Institute of Computer Science and Computer Mathematics,
Jagiellonian University, Cracow, Poland, 
email: {\sl dudajar@gmail.com}
\IEEEcompsocthanksitem
Marcin Paw\l{}owski is with
Institute of Computer Science, Polish Academy of Sciences, Warsaw, Poland,
email: {\sl pawlowski.mp@gmail.com}
\IEEEcompsocthanksitem
Seyit Camtepe is with Data61, CSIRO, Sydney, Australia,
email: {\sl seyit.camtepe@data61.csiro.au}
\IEEEcompsocthanksitem
Arash Mahboubi is with 
School of Computing and Mathematics,
Charles Sturt University,
Port Macquarie, Australia,
email: {\sl amahboubi@csu.edu.au}
\IEEEcompsocthanksitem
Pawe\l{} Morawiecki is with
Institute of Computer Science, Polish Academy of Sciences, Warsaw, Poland,
email: {\sl pawel.morawiecki@gmail.com}
}
}
\maketitle

\begin{abstract}
Compression also known as entropy coding has a rich and long history.
However, a recent explosion of multimedia Internet applications (such as
teleconferencing and video streaming for instance) renews an interest in
fast compression that also squeezes out as much redundancy as possible.
In 2009 Jarek Duda invented his asymmetric numeral system (ANS).
Apart from a beautiful mathematical structure, it is very efficient and offers
compression with a very low residual redundancy.
ANS works well for any symbol source statistics.
Besides, ANS has become a preferred compression algorithm in the IT industry.
However, designing ANS instance requires a random selection of its symbol spread
function. Consequently, each ANS instance offers compression with a slightly different  
compression rate.
\vspace{2mm}

The paper investigates compression optimality of ANS.
It shows that ANS is optimal (i.e. the entropies of encoding and source are equal)
for any symbol sources whose probability distribution is described by natural powers of 1/2.
We use Markov chains to calculate ANS state probabilities. 
This allows us to determine ANS compression rate precisely.
We present two algorithms for finding ANS instances with high compression rates.
The first explores state probability approximations in order to choose ANS instances with
better compression rates.
The second algorithm is a probabilistic one. 
It finds ANS instances, whose compression rate
can be made as close to the best rate as required.
This is done at the expense of the number $\theta$ of internal random ``coin'' tosses.
The algorithm complexity is ${\cal O}(\theta L^3)$, where $L$ is the number of ANS states.
The complexity can be reduced to ${\cal O}(\theta L\log{L})$
if we use a fast matrix inversion.
If the algorithm is implemented on quantum computer, its complexity becomes ${\cal O}(\theta (\log{L})^3)$.

\end{abstract}
\section{Introduction}
An increasing popularity of working from home has dramatically intensified
Internet traffic. To cope with a heavy communication traffic, there are essentially
two options. The first option involves an upgrade of the Internet network.
It is, however, very expensive and not always available.
The second option is much cheaper and employs compression of transmitted symbols.
It also makes sense as typical multimedia communication is highly redundant.
Compression also called {\it entropy coding} has a long history and it can be traced
back to Shannon~\cite{Shannon48} and Huffman \cite{huf1952}.
The well-known Huffman code is the first compression algorithm that works
very well for symbol sources, whose statistics follow natural powers of $1/2$.
Unfortunately, Internet traffic sources almost never have such simple symbol statistics.

Asymmetric numeral systems (ANS) introduced by Duda in \cite{Duda2009} give a very
versatile compression tool.
It allows to compress symbols that occur with an arbitrary probability distribution
(statistics).
ANS is also very fast in both hardware and software.
Currently, ANS gains a significant uptake as a preferred compression algorithm by the IT industry.
It has been adopted by Facebook, Apple, Google, Dropbox, Microsoft and Pixar to name
a few main IT companies (see 
{\sl\blue https://en.wikipedia.org/wiki/Asymmetric\_numeral\_systems}).
ANS can be seen as a finite state machine (FSM) that starts from an initial state,
absorbs symbols from an input source one by one and squeezes out binary encodings.
The sequence of symbols is called a {\it symbol frame}.
The sequence of binary encodings is called a {\it binary frame}.
The heart of an ANS algorithm is its {\it symbol spread function}  (or simply {\it symbol spread} for short)
that assigns FSM states to symbols.
The assignment is completely arbitrary as long as each symbol $s$ is assigned
a number $L_s$ of states such that $p_s\approx L_s/L$, where
$p_s$ is probability of the symbol $s$ and $L=2^R$ is the total number of states
($R$ is a parameter that can be chosen to get an acceptable approximation of $p_s$ by $L_s/L$).

Consequently, there are two closely related problems while designing ANS.
The first also called {\it quantisation} requires from the designer to approximate
a symbol probability distribution ${\cal P}=\{p_s| s\in\mathbb{S}$ by its
approximation ${\cal Q}=\{q_s=\frac{L_s}{L} | s\in\mathbb{S\}}$, where $\mathbb{S}$ is the set of
all symbol (also called symbol source).
It is expected that ANS implemented for  ${\cal Q}$ achieve as close as
possible compression to the source entropy (measured by average encoding length in bits per symbol).
The second problem is selection of a symbol spread for
fixed ANS parameters.
It turns out that some symbol spreads are better than the others.
Again, an obvious goal is to choose them in a such way that average encoding length is as
small as possible and close to the symbol source entropy.
\vspace{2mm}

\noindent
\underline{Motivation.}
Designers of ANS have to strike a balance between efficiency and compression quality.
The first choice that needs to be made is how closely symbol probabilities needs to
be approximated or $p_s\approx \frac{L_s}{L}$.
Clearly, the bigger the number of states ($L=2^R$) the better approximation and better compression.
Unfortunately, a large $L$ slows down compression.
It tuns out that selection of a symbol spread has an impact on the quality of compression.
For some application, this impact cannot be ignored.
This is true, when ANS is applied to build a pseudorandom bit generator.
In this case, the required property of ANS is a minimal residual redundancy, i.e. the average length of
binary frames should be as close as possible to entropy of symbols.
Despite of a growing popularity of ANS, there is no proof of its optimality for
arbitrary symbol statistics.
The work does not aim in proving optimality but rather it develops tools that
allow to compare compression quality of different ANS instances.
It means that we are able to adaptively modify ANS in such way that
every modification provides a compression gain (redundancy reduction).
The following issues are main drivers behind the work:
\vspace{0mm}
\begin{itemize}
\item Investigation of symbol quantisation and its impact on compression quality.
\item Understanding the impact of a chosen symbol spread on the ANS compression quality.
\item Designing an algorithm that allows to build ANS instances that maximise compression
	quality and equivalently minimises the average encoding length.
\end{itemize}
\vspace{1mm}

\noindent
\underline{Contributions.}
We claim the following ones:
\vspace{0mm}
\begin{itemize}
\item Proof of optimality of some ANS instances.
\item Introduction of Markov chains to calculate ANS compression rates.
	Note that Markov chains have been used in the work~\cite{CDMMNPP2022} 
	to analyse ANS-based encryption.
\item Designing ``good''  ANS instances whose state probabilities follow 
	the approximation $\log_2{e}/x$, where $x\in\mathbb{I}$. 
\item A probabilistic algorithm that permits to build ANS, whose compression rate is close or alternatively 
	equal to the best possible. The algorithm uses a pseudorandom number generator (PRNG) as 
	a random ``coin''.
\item An improvement of the Duda-Niemiec ANS cryptosystem that selects at random ANS instances
	with best compression rates.
\end{itemize}

The rest of the work is organised as follows.
Section~\ref{sec_ans} describes ANS and its algorithms.
Section~\ref{sec_optimal} studies the case when ANS produces optimal compression.
Section~\ref{sec_markov_chain} shows how Markov chains can be used to calculate
ANS state equilibrium probabilities and consequently, average lengths of ANS encodings.
Section~\ref{sec_approximation} presents an algorithm that produces ANS instances,
whose state probabilities follow the approximation $\log_2{e}/x$.
Section~\ref{sec_best_ans} describes an algorithm that permits to obtain the best (or close to it)
compression rate. 
Section~\ref{sec_cryptographic} suggests an alternative to the Duda-Niemiec ANS encryption.
The alternative called cryptographic ANS allows to design an ANS secret instance,
whose compression rate is close to the best. 
Section~\ref{sec_experiments} presents results of our experiments and finally
Section~\ref{sec_conclusions} concludes our work.

\section{Asymmetric Numeral Systems}\label{sec_ans}
Here we do not describe the ideas behind ANS design.
Instead, we refer the reader to original papers by Duda \cite{Duda2009,Duda2013} and
the ANS wikipedia page, whose URL address is given in the previous section.
An ANS algorithm suite includes initialisation, compression (symbol frame encoding) and 
decompression (binary frame decoding).
Algorithm \ref{alg_initialisation} shows initialisation.
It also serves as a reference for basic ANS notation.
\begin{algorithm}[h]
 {\scriptsize
 \KwData{a set of symbols $\mathbb{S}$, their probability distribution  $p:\mathbb{S}\to [0,1]$
 and a parameter $R\in\mathbb{N}^+$. }
 \KwResult{ 
  instantiation of encoding and decoding functions:
	\begin{itemize}
	\item $C(s,x)$ and $k_s(x)$;
	\item $D(x)$ and $k(x)$.
	\end{itemize}
	\vspace{0mm}
	}
\KwSteps{proceed as follows:
\vspace{0mm}
	\begin{itemize}
	\item calculate the number of states $L=2^R$;
	\item determine the set of states $\mathbb I=\{L,\ldots,2L-1\}$;
	\item compute integer $L_s\approx Lp_s$, where $p_s$ is probability of $s$ and $s\in\mathbb{S}$;
	\item choose symbol spread function $\overline{s}: {\mathbb I}\to \mathbb{S}$, such that $|\{x\in {\mathbb I}: \overline{s}(x)=s\}|=L_s$;
	\item establish coding function $C(s,y)=x$ for $y\in \{L_s,\ldots,2L_s-1\}$, which assigns states $x\in {\mathbb L_s}$ according to symbol spread function;
	\item compute $k_s(x)=\lfloor \lg(x/L_s) \rfloor$ for $x\in \mathbb I$ and $s\in \mathbb S$. It gives the number of output bits per symbol;
	\item construct decoding function $D(x)=(s,y)$, which for $x\in \mathbb I$, assigns its
		unique symbol (given by the symbol spread function) and the integer $y$,
		where $L_s\leq y\leq 2L_s-1$. Note that $D(x)=C^{-1}(x)$;
  	\item calculate $k(x)=R-\lfloor\lg(x)\rfloor$, which determines
		the number of bitsthat need to be read out from the bitstream;
\end{itemize}
}  
\caption{ANS Initialisation}
 \label{alg_initialisation}
 }
\end{algorithm}
\vspace{2mm}

A compression algorithm accepts a symbol frame $\bf s$ and outputs a binary frame $\bf b$.
In most cases, probability distribution of frame symbols is unknown so it has to be calculated.
This is done by pushing symbols one by one to a stack and counting occurrence of each symbol.
When the last symbol $s_{\ell}$ is processed, the frame statistics is known 
and compression can start.
It means that frame encoding starts from the last symbol $s_{\ell}$.
Algorithm \ref{alg_encoding} describes the ANS compression steps.
{\small 
\begin{algorithm}[h]
 {\scriptsize
 \KwData{ a symbol frame $\mathbf{s}=(s_1,s_2,\ldots, s_{\ell})\in \mathbb{S}^*$ and an initial state $x=x_\ell\in \mathbb{I}$,
 where $\ell=|{\bf s}|$. }
 \KwResult{ a binary frame $\mathbf{b}=(b_1|b_2|\ldots|b_{\ell}) \in \{0,1\}^*$,
	where $|b_i|=k_{s_i}(x_i)$ and $x_i$ is state in $i$-th step.}
\KwSteps{
\For(){$i=\ell,\ell-1,\ldots,2,1$}
{
\vspace{1mm}
 $s:=s_i;$\\
$k=k_s(x)=\lfloor \lg(x/L_s) \rfloor;$ \\
$b_i= x \mod{2^k}$;  \\
$x:=C(s,\lfloor x/2^k\rfloor );$ \\
}
store the final state $x_0=x$\;
}  
\caption{ANS Symbol Frame Encoding}
 \label{alg_encoding}
 }
\end{algorithm}
}
\vspace{0mm}
%
%
%
%
Note that the binary frame is created by concatenation of individual encodings.
This is denoted by $\mathbf{b}=(b_1|b_2|\ldots|b_{\ell})$.
{\small 
\begin{algorithm}[h]
 {\scriptsize
 \KwData{a binary frame $\mathbf{b}\in\{0,1\}^*$ and the final state $x=x_0\in \mathbb{I}$. }
 \KwResult{ symbol frame $\mathbf{s}\in \mathbb{S}^*$.}
\KwSteps{
\While(){$\mathbf{b}\neq \emptyset $ }
{
$(s,y)=D(x)$; \\
$k=k(x)=R-\lfloor\lg(x)\rfloor$; \\
$b=MSB(\mathbf{b})_k$; \\
$\mathbf{b}:= LSB(\mathbf{b})_{|\mathbf{b}|-k}$; \\
$x:=2^k y +b$; \\
}
}  
\caption{ANS Binary Frame Decoding}
 \label{alg_decoding}
 }
\end{algorithm}
}
\vspace{0mm}
Decompression steps are shown in Algorithm \ref{alg_decoding}.
Note that $LSB(\mathbf{b})_{\ell}$ and $MSB(\mathbf{b})_{\ell}$ stand for the $\ell$ least
and most significant bits of $\mathbf{b}$, respectively.

\subsection{ANS Example}
Given a symbol source $\mathbb{S}=\{s_1,s_2, s_3 \}$,
where $p_1=\frac{3}{16}$, $p_2=\frac{5}{16}$, $p_3=\frac{8}{16}$ and the parameter ${R}=4$.
The number of states is $L=2^R=16$ and the state set equals to $\mathbb I=\{16,17,\ldots , 31\}$.
A symbol spread $\overline{s}: {\mathbb I}\to \mathbb{S}$ is assumed to be
{\small
\[
		\overline{s}(x)=
	\left\{
	\begin{array}{cl}
	  s_1 &  \mbox{ if } x\in\{18,22,25\} =\mathbb L_1 \\
	  s_2 &  \mbox{ if }  x \in \{19,20,23,26,28\}=\mathbb L_2\\
	  s_3 & \mbox{ if } x\in\{16,17,21,24,27,29,30,31\}=\mathbb L_3
	\end{array}
	\right.  \;\; \longleftrightarrow \;\;\; 
\left.
\arraycolsep=0.6pt\def\arraystretch{1.4}
\begin{array}{|cccc cccc cccc cccc|} \hline
16 & 17 & 18 & 19 & 20 & 21 & 22 & 23 & 24 & 25 & 26 & 27 & 28 & 29 & 30 & 31 \\ \hline
 3 & 3 & 1 & 2 & 2 & 3 & 1 & 2 & 3 & 1 & 2 & 3 & 2 & 3 & 3 & 3 \\ \hline
\end{array}
\right.,
\]
}
where $L_1=3$, $L_2=5$ and
$L_3=8$.
The frame encoding table 
${\mathbb E}(x_i,s_i)=({x_{i+1}, b_i})\stackrel{\mbox{\it \scriptsize def}}{\equiv} {x_{i+1}\choose b_i}$ 
is shown in Table \ref{tab_example_1}.
\vspace{-2mm}
\begin{table}[h]
\caption{ANS for $16$ states (symbol probabilities $\{3/16$, 5/16$, 8/16\}$)}
\vspace{-8mm}
\begin{center}
	\[
	\def\arraystretch{1.5}
	\begin{array}{| c || c|c|c|c |c|c|c|c| c|c|c|c|  c|c|c|c|} \hline
	s_i \backslash x_i &16&17&18&  19&20&21&22&  23&24&25&26& 27&28&29&30&31 \\ \hline
	s_1&{22\choose 00}&{22\choose 01}&{22\choose 10}&{22\choose 11}  &
	{25\choose 00}&{25\choose 01}&{25\choose 10}&{25\choose 11}  &
	{18\choose 000}&{18\choose 001}&{18\choose 010}&{18\choose 011} &{18\choose 100}&
	{18\choose 101}&{18\choose 110}& {18\choose 111}\\ \hline
	s_2&{26\choose 0}&{26\choose 1}&{28\choose 0}&{28\choose 1}  &
	{19\choose 00}&{19\choose 01}&{19\choose 10}&{19\choose 11}  &
	{20\choose 00}&{20\choose 01}&{20\choose 10}&{20\choose 11} &
	{23\choose 00}&{23\choose 01}&{23\choose 10}& {23\choose 11} \\ \hline
	 s_3&{16\choose 0}&{16\choose 1}&{17\choose 0}&{17\choose 1}  &{21\choose 0}&{21\choose 1}&
	{24\choose 0}&{24\choose 1}  &{27\choose 0}&{27\choose 1}&{29\choose 0}&{29\choose 1} &
	{30\choose 0}&{30\choose 1}&{31\choose 0}&{31\choose 1} \\ \hline
	\end{array}
	\]
\end{center}
\label{tab_example_1}
\end{table}%
The example is used throughout the work to illustrate our considerations.

\vspace{-2mm}
\section{Optimal ANS}\label{sec_optimal}
Given an ANS instance designed for a symbol source with its probability distribution
$p=\{p_s; s\in\mathbb{S}\}$ and the parameter $R$. The set of all ANS states is
$\mathbb{I}=\{2^R,\ldots, 2^{R+1}-1\}$.
ANS is optimal if the average length of binary encoding is equal to the symbol entropy or
\[
	\frac{1}{\ell}\sum_{i=1}^{\ell} |b_i|=H(\mathbb{S})=\sum_{s\in \mathbb{S}} p_s\log_2{p_s^{-1}},
\]
where a symbol frame  $\mathbf{s}=(s_1,s_2,\ldots, s_{\ell})$ consists of $\ell$ symbols.
The same symbols of the frame can be grouped so we get
\[
	\frac{1}{\ell}\sum_{i=1}^{\ell} |b_i| = \frac{1}{\ell} \sum_{s\in\mathbb{S}} \sum_{s\in\mathbf{s}} |b_s| =
	 \sum_{s\in\mathbb{S}}\frac{\sum_{s\in\mathbf{s}} |b_s|}{\ell}.
\]
Note that $\frac{1}{\ell} \sum_{s\in\mathbf{s}} |b_s|=p_s\log_2{p_s^{-1}}$. 
So we have proven the following lemma.
\begin{lemma}\label{lemma_optimal}
Given a symbol source $\mathbb S$ whose probability distribution is $p=\{p_s; s\in\mathbb{S}\}$
and ANS for $\mathbb S$.
Then ANS is optimal if and only if 
\[
	\frac{1}{\ell} \sum_{s\in\mathbf{s}} |b_s|=p_s\log_2{p_s^{-1}}\mbox{ for all } s\in\mathbb{S},
\]
where $\mathbf{s}$ is a symbol frame, which repeats every symbol $\ell \cdot p_s$ times.
\end{lemma}

Let us consider how ANS encodes symbols.
According to Algorithm~\ref{alg_encoding}, a symbol $s$ is encoded into $b_i=x\pmod{2^k}$,
where $k_s=\lfloor \log_2(x/L_s) \rfloor$ and $x\in\{2^R,\ldots,2^{R+1}-1\}$.
The length $k_s$ of encoding depends on the state $x$.
The shortest encoding is when $x=2^R$ and the longest when $x=2^{R+1}-1$.
It means that $\lfloor \log_2(2^R/L_s) \rfloor \leq k_s \leq \lfloor \log_2((2^{R+1}-1)/L_s) \rfloor$.
As $L_s=2^R p_s$, we get
$\lfloor \log_2p_s^{-1} \rfloor \leq k_s \leq \lfloor \log_2p_s^{-1}(2-\frac{1}{2^R}) \rfloor$.
Consider a general case when $2^{-(i+1)} < p_s < 2^{-i}$.
Then $k_s\in\{i, i+1\}$.
If $p_s=2^{-i}$, then 
the inequality $\lfloor \log_22^i \rfloor \leq k_s \leq \lfloor \log_2 2^i(2-\frac{1}{2^R}) \rfloor$
points to a single encoding length $k_s=i$.
The above leads us to the following conclusion.
\begin{lemma}\label{lemma_encoding_lenght}
Given ANS described by Algorithm \ref{alg_encoding}. Then a symbol $s$ is encoded into a binary
string of the length $k_s$, where
\begin{itemize}
\item $k_s=i$ if $p_s=2^{-i}$,
\item $k_s\in\{i, i+1\}$ if $2^{-(i+1)} < p_s < 2^{-i}$. This includes an interesting
	case when $2^{-1} < p_s < 2^{0}$ with $k_s\in\{0,1\}$ -- some symbols
	are encoded into void bits $\varnothing$.
\end{itemize}
\end{lemma}
\noindent
The above lemmas lead us to the following conclusions.
\begin{itemize}
\item ANS provides optimal encoding for a symbol, whose probability is a natural power of $1/2$.
\item Optimal ANS exists for symbol sources, whose all symbol probabilities are natural powers of $1/2$.
\item ANS and the Huffman code share compression optimality. Unlike the Huffman code,
	ANS compresses well also symbol with an arbitrary probability distribution.
\end{itemize}
Table~\ref{tab_example_1} shows encoding for three symbols.
One of them occurs with probability $p_{s_1}=1/2$.
Note that ANS assigns a single bit encoding for all ANS states.
This illustrates optimal encoding of $s_1$.

\section{State Probabilities and Markov Chains}\label{sec_markov_chain}
ANS can be looked at as FSM, whose internal state $x\in\mathbb{I}$ changes during compression.
Behaviour of ANS states is probabilistic and 
can be characterised by state probability distribution $\{p_x; x\in\mathbb{I}\}$.
For a given symbol $s\in\mathbb{S}$, the average encoding length of $s$ is
$\kappa(s)=\sum_{x\in\mathbb{I}} k_s(x) p_x$,
where $k_s(x)$ is the length of an encoding assigned to $s$ when ANS is in the state $x$
(see Algorithm~\ref{alg_encoding}).
The average length of ANS encodings is $\kappa=\sum_{s\in\mathbb{S}}p_s\kappa(s)$.
When we deal with an optimal ANS, then $\kappa=H(\mathbb{S})$, which also means
that $\kappa(s)=H(s)=\log_2{p_s^{-1}}$ for all $s\in\mathbb{S}$.
Typically, compression quality is characterised by a ratio between the length of symbol frame
and the length of binary frame.
A better measure from our point of view is a residual redundancy per symbol,
which is defined as $\Delta H=\kappa-H(\mathbb{S})$. 
The measure has the following advantages:
\vspace{-0mm}
\begin{itemize}
\item easy identification of an optimal ANS instance when its $\Delta H=0$;
\item quick comparison of two ANS instances -- a better ANS has a smaller $\Delta H$;
\item fast calculation of the length of a redundant part of binary frame, which is $\ell\cdot \Delta H$ bits,
	where $\ell$ is the number of symbols in the input frame.
\end{itemize}

To determine $\Delta H$ for an ANS instance, it is necessary to calculate 
probability distribution $\{p_x; x\in\mathbb{I}\}$.
It depends on a (random) selection of symbol spread.
Fortunately, ANS state transition can be modelled as a Markov chain~\cite{PB2020,hag2002}.
It is reasonable to expect that probabilities of ANS states of their Markov chain 
attain an equilibrium after processing a sufficiently long sequence of symbols.
Note that the ANS Markov chain is in equilibrium when state probabilities do not change
after processing single symbols.
{\small 
\begin{algorithm}[h]
 {\scriptsize
 \KwData{ANS encoding table $\mathbb{E}(s,x)$ and symbol probability distribution $\{p_s; s\in\mathbb{S}\}$.}
 \KwResult{ probability distribution ${\cal P}_{eq}=\{p_x; x\in\mathbb{I}\}$ of ANS Markov chain equilibrium.}
\KwSteps{
 \begin{itemize}
\item initialise $2^R\times 2^R$ matrix $M$ to all zeros except the main diagonal 
	$M[{i,i}]=-1$ for $i=0,\ldots,2^R-1$ and the last row, where $M[2^R-1,j]=1$ for $j=1,\ldots,2^R$; 
\item create a vector $B$ with $2^R$ entries with all zero entries except 
	the last entry equal to $1$;
 \end{itemize}
\For{$x\in\{2^R,\ldots,2^{R+1}-2\}$ and $s\in\mathbb{S}$} 
{
find $\mathbb{E}(s,x)=x'$; \\
$M[x' \!\!\!\!\mod{2^R},x  \!\!\!\!\ \mod{2^R}]:= M[x' \!\!\! \mod{2^R},x  \!\!\! \mod{2^R}] + p_s$; 
}
$\bullet$ solve $M\cdot X=B$ using Gaussian elimination; \\
$\bullet$ return  $X$ $\rightarrow$ ${\cal P}_{eq}=\{p_x; x\in\mathbb{I}\}$;
}  
\caption{Equilibrium of ANS States}
 \label{alg_markov}
 }
\end{algorithm}
}
\vspace{0mm}
%
%
%
%
Algorithm~\ref{alg_markov} shows steps for construction of a system of linear equations,
whose solution gives equilibrium probability distribution
${\cal P}_{eq}=\{p_x; x\in\mathbb{I}\}$.
The encoding table $\mathbb{E}(s,x)=x'$ shows transition of a state $x$ to the next state $x'$
while encoding/compressing a symbol $s$. 
\begin{example}\label{example_markov}
Consider the ANS instance from Table~\ref{tab_example_1}.
According to the above algorithm, we get a matrix $M$ as follows:
\vspace{-4mm}
\begin{center}
%
{\scriptsize
\[
M=
\left[
\arraycolsep=0.6pt\def\arraystretch{1.4}
\begin{array}{cccc cccc cccc cccc}
- \frac{1}{2} &  \frac{1}{2} &     0 &     0 &      0 &      0 &      0 &      0 &      0 &      0 &       0 &     0 &      0 &      0 &       0 &     0 \\
 0 &     -1 &      \frac{1}{2} &   \frac{1}{2} &    0 &      0 &      0 &      0 &      0 &      0 &       0 &     0 &      0 &      0 &       0 &     0 \\
 0 &      0 &      -1 &    0 &      0 &      0 &      0 &      0 &      \frac{3}{16} &  \frac{3}{16} &  \frac{3}{16} &  \frac{3}{16} &  \frac{3}{16} &  \frac{3}{16} &  \frac{3}{16} &  \frac{3}{16}\\
 0 &      0 &      0 &     -1 &  \frac{5}{16} &  \frac{5}{16} &  \frac{5}{16} &  \frac{5}{16} &     0 &      0 &       0 &     0 &      0 &      0 &       0 &     0\\
 0 &      0 &      0 &      0 &     -1 &     0 &      0 &      0 &     \frac{5}{16} &  \frac{5}{16} &  \frac{5}{16} &  \frac{5}{16} &   0 &      0 &       0 &     0\\
 0 &      0 &      0 &      0 &     \frac{1}{2} &     - \frac{1}{2} &  0 &     0 &      0 &      0 &       0 &     0 &      0 &      0 &       0 &     0\\
  \frac{3}{16} & \frac{3}{16} &  \frac{3}{16} &  \frac{3}{16} &  0 &      0 &     -1 &     0 &      0 &      0 &       0 &     0 &      0 &      0 &       0 &     0\\
 0 &      0 &      0 &       0 &     0 &      0 &      0 &    -1 &      0 &      0 &       0 &     0 &    \frac{5}{16} &  \frac{5}{16} &  \frac{5}{16} &  \frac{5}{16}\\
 0 &      0 &      0 &       0 &     0 &     0 &       \frac{1}{2} &  \frac{1}{2} &     -1 &     0 &       0 &     0 &      0 &      0 &       0 &     0\\
 0 &      0 &      0 &       0 &   \frac{3}{16} & \frac{3}{16} &  \frac{3}{16} &  \frac{3}{16} &     0 &    -1 &       0 &     0 &      0 &      0 &       0 &     0\\
  \frac{5}{16} & \frac{5}{16} &  0 &     0 &      0 &      0 &       0 &      0 &      0 &      0 &      -1 &    0 &      0 &      0 &       0 &     0\\
 0 &      0 &      0 &       0 &     0 &      0 &       0 &      0 &     \frac{1}{2} &  \frac{1}{2} &      0 &    -1 &      0 &      0 &       0 &     0\\
 0 &      0 &    \frac{5}{16} &   \frac{5}{16} &   0 &      0 &      0 &      0 &       0 &     0 &      0 &      0 &    -1 &      0 &       0 &     0\\
 0 &     0 &      0 &      0 &       0 &      0 &      0 &      0 &       0 &     0 &      \frac{1}{2} &  \frac{1}{2} &      0 &     -1 &      0 &     0\\
 0 &     0 &      0 &      0 &       0 &      0 &      0 &      0 &       0 &     0 &      0 &     0 &     \frac{1}{2} &     \frac{1}{2} &     -1 &    0\\
1 &	1 &	1 &	1 & 	1 &	1 &	1 &	1 &	 1 &	1 &	1 &	1 & 	1 &	1 &	1 &	1
 \end{array}
\right]
\]
}
\end{center}
The vector $B=[0,0,0,0, 0,0,0,0, 0,0,0,0, 0,0,0,1]$.
The equilibrium probabilities $(p_{16},\ldots,p_{31})$ are 
{\footnotesize
\begin{eqnarray*}
\left( \frac{367}{4590},\frac{367}{4590},\frac{1933}{24480},\frac{1189}{14688},\frac{991}{14688},\frac{991}{14688},\frac{367}{6120},\frac{157}{2448}\right.,\\
\left. \frac{1519}{24480}, \frac{1189}{24480}, \frac{367}{7344}, \frac{677}{12240}, \frac{367}{7344}, \frac{1933}{36720}, \frac{157}{3060}, \frac{157}{3060}\right).
\end{eqnarray*}}
Now it is easy to calculate both the average encoding length $\kappa= 1.4790168845$ and symbol entropy
$H(\mathbb{S})=1.4772170014$.
ANS leaves a residual redundancy $\Delta H=\kappa-H(\mathbb{S})=0.0017998831$ bits per symbol.
\vspace{2mm}

\noindent
Let us change the symbol spread by swapping states $25$ with $28$ so
we have the following symbol spread
{\small
\[
		\overline{s}(x)=
%
\left.
\arraycolsep=0.6pt\def\arraystretch{1.4}
\begin{array}{|c|c|c|c| c|c|c|c| c|c|c|c| c|c|c|c|} \hline
16 & 17 & 18 & 19 & 20 & 21 & 22 & 23 & 24 & 25 & 26 & 27 & 28 & 29 & 30 & 31 \\ \hline
 3 & 3 & 1 & 2 &    2 & 3 & 1 & 2 &     3 & 2 & 2 & 3 &     1 & 3 & 3 & 3 \\ \hline
\end{array}
\right..
\]}
The encoding function $C(s,y)$ is:
{\small
\[\arraycolsep=1.4pt
	\begin{array}{| c || c|c|c|c |c|c|c|c|  c|c|c|c|  c|} \hline
	 s \backslash y &3&4&5& 6 &7&8&9& 10 &11&12&13& 14 &15 \\ \hline
	 s_1 &18&22&28& - &-&-&-& - &-&-&-& - & -\\ \hline
	 s_2 &-&-&19& 20 &23&25&26& - &-&-&-& - & -\\ \hline
	 s_3 &-&-&-& - &-&16&17& 21 &24&27&29& 30 & 31\\ 
	 \hline
	\end{array}
\]}
The encoding table is given in Table \ref{tab_example_2}.
\begin{table}[h]
\caption{ANS for $16$ states with state swap (symbol probabilities $\{3/16, 5/16, 8/16\}$)}
\vspace{-10mm}
\begin{center}
	\[
	\def\arraystretch{1.5}
	\begin{array}{| c || c|c|c|c |c|c|c|c| c|c|c|c|  c|c|c|c|} \hline
	s_i \backslash x_i &16&17&18&  19&20&21&22&  23&24&25&26& 27&28&29&30&31 \\ \hline
	s_1&{22\choose 00}&{22\choose 01}&{22\choose 10}&{22\choose 11}  &
	{28\choose 00}&{28\choose 01}&{28\choose 10}&{28\choose 11}  &
	{18\choose 000}&{18\choose 001}&{18\choose 010}&{18\choose 011} &{18\choose 100}&
	{18\choose 101}&{18\choose 110}& {18\choose 111}\\ \hline
	s_2&{25\choose 0}&{25\choose 1}&{26\choose 0}&{26\choose 1}  &
	{19\choose 00}&{19\choose 01}&{19\choose 10}&{19\choose 11}  &
	{20\choose 00}&{20\choose 01}&{20\choose 10}&{20\choose 11} &
	{23\choose 00}&{23\choose 01}&{23\choose 10}& {23\choose 11} \\ \hline		
	s_3&{16\choose 0}&{16\choose 1}&{17\choose 0}&{17\choose 1}  &{21\choose 0}&{21\choose 1}&
	{24\choose 0}&{24\choose 1}  &{27\choose 0}&{27\choose 1}&{29\choose 0}&{29\choose 1} &
	{30\choose 0}&{30\choose 1}&{31\choose 0}&{31\choose 1} \\ \hline
	\end{array}
	\]
\end{center}
\vspace{-2mm}
\label{tab_example_2}
\end{table}%

After running Algorithm~\ref{alg_markov}, we get equilibrium probabilities $(p_{16},\ldots,p_{31})$ as follows:
{\footnotesize
\begin{eqnarray*}
\left( \frac{3071}{38400}, \frac{3071}{38400}, \frac{8077}{102400}, \frac{4981}{61440}, \frac{4177}{61440},  \frac{4177}{61440},  \frac{3071}{51200},  \frac{65}{1024}, \right.\\
\left.  \frac{6321}{102400},  \frac{3071}{61440},  \frac{3071}{61440},  \frac{17159}{307200},  \frac{4981}{102400},  \frac{5419}{102400},  \frac{13}{256},  \frac{13}{256}\right).
 \end{eqnarray*}
}
The average encoding length $\kappa=1.4789314193$.
This ANS instance leaves residual redundancy $\Delta H=\kappa-H(\mathbb{S})=0.0017144179$ bits per symbol.
Clearly, it offers better compression than the first variant. \hfill $\Box$
\end{example}

The following remarks summarise the above discussion:
\begin{itemize}
\item ANS offers a very close to optimal compression, when the parameter $R$ is big enough so $L_s=2^R p_s$
	for all $s\in\mathbb{S}$. In general, however, for arbitrary symbol statistics, ANS tends to leave a residual
	redundancy $\Delta H\neq 0$.
\item In some applications, where ANS compression is being used heavily (for instance, video/teleconference streaming),
	it makes sense to optimise ANS so $\Delta H$ is as small as possible.
	Note that the smaller $\Delta H$ the more random are binary frames. This may increase security level of
	systems that use compression (for instance, joint compression and encryption).
\item It is possible to find the best ANS instance by exhaustive search through all distinct symbol spreads.
	For $L=2^R$ states, we need to search through $\frac{L!}{\prod_{s\in\mathbb{S}}L_s!}$ 
	ANS instances.
	This is doable for a relatively small number of states. For ANS in our example,
	the search space includes $\frac{16!}{8!\cdot 5!\cdot 3!}=720720$ instances.
	For a bigger $L$ (say, above 50), the exhaustive search is intractable.
\end{itemize}

\section{Tuning ANS Symbol Spreads}\label{sec_approximation}
 For the compression algorithms such as Zstandard and LZFSE, a typical symbol source 
contains $n=256$ elements. 
To get a meaningful approximation of the source statistics, we need a bigger number $L$ of states.
A rough evaluation of compression quality penalty given in \cite{Duda2021} tells us that
choosing $L=2n$ incurrs a entropy loss of $\Delta  H=0.01$ bits/symbol;
if $L=4n$, then $\Delta H=0.003$ and if $L=8n$, then $\Delta H=0.0006$.
This confirms an obvious intuition that the bigger number of states $L$ the better
approximation and consequently compression.
However, there is a ``sweet spot'' for $L$, where its further increase 
slows compression algorithm but without a noticeable compression rate gain.

Apart from approximation accuracy of symbol probabilities so $p_s\approx L_s/L$, symbol spread
$\overline{s}: {\mathbb I}\to \mathbb{S}$ has an impact on compression rate.
Intuitively, one would expect that symbol spread is chosen uniformly at random from all
$\frac{L!}{\prod_{s\in\mathbb{S}} L_s!}$ possibilities. 
It is easy to notice that they grow exponentially with $L$.
An important observation is that probability distribution of ANS states during compression is not uniform.
In fact, a state $x\in {\mathbb I}$ occurs with probability that can be approximated as
$p_x\approx \log_2(e)/x$.
Note that this is beneficial for a compression rate as smaller states (with shorter encodings) are preferred over
larger ones (with longer encodings).
The natural probability bias of states has an impact on compression rate making some symbol spreads
better than the others.
Let us take a closer look at how to choose symbol spread so it maintains the natural bias.

Recall that for  a given symbol $s\in\mathbb{S}$ and state $x\in\mathbb{I}$,
the encoding algorithm (see Algorithm~\ref{alg_encoding})
calculates $k=k_s=\lfloor \log_2 \frac{x}{L_s} \rfloor$; extracts $k$ least significant bits of the state as the encoding of $s$ and
finds the next state $x'=C(s,\lfloor \frac{x}{2^k}\rfloor)$, where $C(s,y)$ is equivalent to a symbol spread $\overline{s}$
and $x'\in\mathbb{L}_s$.
Consider properties of coding function $C(s,y)$ that are used in our discussion.
\begin{fact}\label{fact_intervals}
Given a symbol $s$ and coding function $C(s,\lfloor \frac{x}{2^k}\rfloor)$.
Then the collection of states $\mathbb{I}$ is divided into $L_s$ state intervals $\mathbb{I}_i$, where
\vspace{-0mm}
\begin{itemize}
\item each interval $\mathbb{I}_i$ consists of all consecutive states that 
	share the same value $\lfloor \frac{x}{2^k}\rfloor$ for all $x\in\mathbb{I}_i$,
\item coding function assigns the same state $x'=C(s,\lfloor \frac{x}{2^k}\rfloor)$ for $x\in\mathbb{I}_i$ and $x'\in \mathbb{L}_s$,
\item the cardinality of $\mathbb{I}_i$ is $2^k$, where $k=k_s=\lfloor \log_2 \frac{x}{L_s} \rfloor$ and
	 $\mathbb{I}=\cup_{i=1,\ldots,L_s} \mathbb{I}_i$.
\end{itemize}
\end{fact}
\begin{example}
Take into account ANS described in Table~\ref{tab_example_1}. For the symbol $s_1$, 
the collection of states $\mathbb{I}=\{16,\ldots,31\}$ splits into $L_{s_1}=3$ intervals as follows
\begin{center}
\arraycolsep=1.6pt\def\arraystretch{1.4}
\begin{tabular}{|c|c|c|} \hline
$\;\;\mathbb{I}_1=\{16,17,18,19\}$ & $\;\;\mathbb{I}_2=\{20,21,22,23\}$ & $\;\;\mathbb{I}_3=\{24,25,26,27,28,29,30,31\}$ \\
\hline
$C(s,\lfloor \frac{x}{4}\rfloor)=22$; $x\in\mathbb{I}_1$  & $C(s,\lfloor \frac{x}{4}\rfloor)=25$; $x\in\mathbb{I}_2$ 
& $C(s,\lfloor \frac{x}{8}\rfloor)=18$; $x\in\mathbb{I}_3$ \\
\hline
\end{tabular}
\end{center}
Note that cardinalities of $\mathbb{I}_i$ are powers of 2 or $|\mathbb{I}_1|=|\mathbb{I}_2|=4$ and $|\mathbb{I}_3|=8$.
As the encoding function $C(s,\cdot)$ is constant in the interval $\mathbb{I}_i$, it makes sense to use a shorthand
$C(s,\mathbb{I}_i)$ instead of $C(s,\lfloor \frac{x}{2^k}\rfloor)$ for $x\in\mathbb{I}_i$.
$\;\;$ \hfill $\Box$
\end{example}
Assume that we know probabilities $p(x)$ of states of $x\in \mathbb{I}$,
then the following conclusion can be derived from Fact~\ref{fact_intervals}.
\begin{corollary}
Given an ANS instance defined by Algorithms \ref{alg_initialisation}, \ref{alg_encoding} and \ref{alg_decoding}.
Then for each symbol $s\in\mathbb{S}$, state probabilities have to satisfy the following relations:
\begin{equation}\label{eq_prob_relation}
	p_s \sum_{x\in \mathbb{I}_i} p(x) =p( C(s,\mathbb{I}_i));\;\;\;  i=1,\ldots, L_s.
\end{equation}
\end{corollary}
Recall that state probabilities can be approximated by $p(x)\approx \log_2(e)/x$.
As Equation~\ref{eq_prob_relation} requires summing up probabilities of consecutive states,
we need the following fact.
\begin{fact}
Given an initial part of the harmonic series $(1+\frac{1}{2} +\ldots + \frac{1}{r})$,
then it can approximated as shown below
\begin{equation}\label{eq_harmonic_approx}
\sum_{i=1}^{r} \frac{1}{i} =1+\frac{1}{2} +\ldots + \frac{1}{r} \approx \ln(r)+\gamma,
\end{equation}
where the constant $\gamma\approx 0.577$.
It is easy to get approximation of the series $(\frac{1}{r}+\ldots+\frac{1}{r+\alpha})$, which is
$\approx \ln(r+\alpha) -\ln(r-1)=\ln (\frac{r+\alpha}{r-1})$, where $\alpha\in\mathbb{N}^+$.
\end{fact} 
Now, we ready to find out a preferred state for our symbol spread.
Let us take into account Equation~\ref{eq_prob_relation}.
Using the above established facts and our assumed approximation $p(x)\approx \log_2(e)/x$, 
the left-hand side of the equation becomes
\[
	p_s \sum_{x\in \mathbb{I}_i} p(x)=p_s \log_2(e) \sum_{x\in \mathbb{I}_i} \frac{1}{x}=
	p_s \log_2(e)\ln \frac{r+\alpha-1}{r-1},
\]
where $r$ is the first state in $\mathbb{I}_i$ and $(r+\alpha-1)$ -- the last and $\alpha=2^k$.
As we have assumed that the state $C(s,\mathbb{I}_i)\approx \log_2(e) \frac{1}{y}$,
where $y$ points the preferred state that needs to be included into $\mathbb{L}_s$, we get
\[
	p_s \log_2(e)\ln \frac{r+\alpha-1}{r-1} =  \log_2(e) \frac{1}{y}.
\]
This brings us to the following conclusion.
\begin{corollary}
Given ANS as defined above, then a preferred state $y$ for $\mathbb{I}_i$ determined for a symbol $s$ is
\begin{equation}\label{eq_preferred}
	y= \left( p_s \ln\frac{r+\alpha-1}{r-1} \right)^{-1},
\end{equation}
where $\mathbb{I}_i= [ r,\ldots,r+\alpha-1 ]$ and $\lfloor y\rceil$ an integer closest to $y$ and it is added to $\mathbb{L}_s$.
\end{corollary}
{\small 
\begin{algorithm}[h]
 {\footnotesize
 \KwData{ANS number of states $L=2^R$ and symbol probability distribution $\{p_s; s\in\mathbb{S}\}$.}
 \KwResult{ symbol spread $\overline{s}$ determined by $\mathbb{L}_s$ for $s\in\mathbb{S}$.}
 \KwSteps{
 initialise $\mathbb{L}_s=\varnothing$ for $s\in\mathbb{S}$; \\
 \For(){$s\in\mathbb{S}$}
{
 $\bullet$ compute $L_s=L\cdot p_s$;  \\
 $\bullet$ split $\mathbb{I}$ into $\mathbb{I}_i$; where $|\mathbb{I}_i|=\lfloor \log_2 \frac{x}{L_s} \rfloor$, $x\in\mathbb{I}_i$ and 
 	$\;\;$ $i=1,\ldots L_s$; \\
	\For(){ $i=1,\ldots, L_s$ }
	{
	 $\bullet$ find $y= \left( p_s \ln\frac{r+\alpha-1}{r-1} \right)^{-1}$, 
	 $\;\;\;$where $ \mathbb{I}_i= [r,\ldots,r+\alpha-1 ]$;\\
	 $\bullet$ $\mathbb{L}_s:=\mathbb{L}_s \cup \lfloor y \rceil $;
	}
}
$\bullet$ remove collisions from $\mathbb{L}_s$ so $\mathbb{L}_s \cap \mathbb{L}_{s'}=\varnothing$ for $s\neq s'$; \\
$\bullet$ return $\overline{s}$ or equivalently $\mathbb{L}_s$ for $s\in\mathbb{S}$;
}  
\caption{Symbol Spread Tuning}
 \label{alg_symbol_spread_1}
 }
\end{algorithm}
}
\vspace{0mm}
%
%
%
%
%
Algorithm~\ref{alg_symbol_spread_1} shows an algorithm for calculation of symbol spread
with preferred states.
Let us consider the following example.
\begin{example}
Take ANS from Section~\ref{sec_ans} with $L=16$ states and three symbols $\mathbb{S}=\{s_1,s_2,s_3\}$ that occur with
probabilities $3/16$, $5/16$ and $8/16$, respectively.
For $s_1$, the set of states $\mathbb{I}$ splits into $L_{s_1}=16 p_{s_1}=3$ subsets:
$\mathbb{I}_1=\{16,17,18,19\}$, $\mathbb{I}_2=\{20,21,22,23\}$ and $\mathbb{I}_3=\{24,\ldots,31\}$.
Let us compute the preferred state $y$ for $\mathbb{I}_1$.
It is $y= \left( p_s \ln\frac{r+\alpha-1}{r-1} \right)^{-1}\approx 22.56$, where $r=16$ and $r+\alpha-1=19$.
For $\mathbb{I}_2$, we get $y\approx 27.91$.
For $\mathbb{I}_3$, we obtain $y\approx 17.86$.
After rounding to closest integers, $\mathbb{L}_{s_1}=\{18,23,28\}$.
In similar way we calculate $\mathbb{L}_{s_2}=\{17,20,23,26,29\}$ and
$\mathbb{L}_{s_3}=\{16,18,20,22,24,26,28,30\}$.
Clearly, there are few collisions, for example state $18$ belongs to both $\mathbb{L}_{s_1}$ and $\mathbb{L}_{s_3}$.
They need to be removed. We accept $\mathbb{L}_{s_3}$. The colliding states in other sets are replaced by their
closest neighbours, which are free. This obviously makes sense as neighbour states share similar probabilities.
The final symbol spread is as follows
\[\small
\arraycolsep=1.6pt\def\arraystretch{1.4}
\begin{array}{|c|c|c|c| c|c|c|c| c|c|c|c| c|c|c|c|} \hline
16 & 17 & 18 & 19 & 20 & 21 & 22 & 23 & 24 & 25 & 26 & 27 & 28 & 29 & 30 & 31 \\ \hline
3&2&3&1& 3&2&3&2& 3&1&3&2& 3&2&3&1 \\ \hline
\end{array}
\vspace{2mm}
\]
We can compute the average encoding length by computing equilibrium probabilities, which for the above
symbol spread is $\kappa=\frac{3619}{2448}\approx 1.4783$. It turns out that this is the best compression rate as argued
in the next Section.
In contrast, average lengths of symbol spreads in Example~\ref{example_markov} are $1.4790$ and $1.4789$.
\hfill $\Box$
\end{example}

\noindent
Algorithm~\ref{alg_symbol_spread_1} may produce many symbol spreads
with slightly different compression rates.
Preferred positions need to be rounded to the closest integers.
Besides, there may be collisions in sets $\mathbb{L}_s$ ($s\in\mathbb{S}$) that have to be removed.
Intuitively, we are searching for a symbol spread that is the best match for preferred positions.
In other words, we need to introduce an appropriate distance to measure the match.
\begin{definition}
Given ANS with a symbol spread $\mathbb{L}_{s\in\mathbb{S}}$ and
a collection of preferred positions ${\cal L}_{s\in\mathbb{S}}$
calculated according to Equation~\ref{eq_preferred}.
Then the distance between them is computed as
\begin{equation}\label{eq_distance}
d(\mathbb{L}_{s\in\mathbb{S}}, {\cal L}_{s\in\mathbb{S}})=
\sum_{s\in\mathbb{S}} \sum_{x\in\mathbb{L}_s}{|  x - y  |},
\end{equation}
where $y$ is the preferred position that is taken by $x$.
\end{definition}
Finding the best match for a calculated ${\cal L}_{s\in\mathbb{S}}$ is equivalent to 
identification of a symbols spread $\mathbb{L}_{s\in\mathbb{S}}^*$
such that 
\begin{equation}\label{eq_optimal}
d(\mathbb{L}_{s\in\mathbb{S}}^*, {\cal L}_{s\in\mathbb{S}})= \min_{\mathbb{L}_{s\in\mathbb{S}}}
d(\mathbb{L}_{s\in\mathbb{S}}, {\cal L}_{s\in\mathbb{S}}),
\end{equation}
where the symbol spread $\mathbb{L}_{s\in\mathbb{S}}$ runs through all possibilities.
{\small 
\begin{algorithm}[h]
 {\footnotesize
 \KwData{ANS number of states $L=2^R$, symbol probability distribution $\{p_s; s\in\mathbb{S}\}$ and $ {\cal L}_{s\in\mathbb{S}}$}
 \KwResult{ symbol spread $\overline{s}$ determined by $\mathbb{L}_s^*$ for $s\in\mathbb{S}$}
 \KwSteps{\\
 \begin{itemize}
\item create a table with three rows and $L=2^R$ columns;\\
\item put all consecutive states (i.e. $(L,L+1,\ldots, 2L-1)$ in increasing order in the first row; \\
\item insert all numbers from  ${\cal L}_{s\in\mathbb{S}}$ in increasing order in the second row together with their symbol labels in the third row; \\
\item read out all states from the first row that correspond to appropriate symbol labels (in the third row);\\
\item return $\overline{s}$ or equivalently $\mathbb{L}_s^*$ for $s\in\mathbb{S}$;
 \end{itemize}
}  
\caption{Finding Symbol Spread $\mathbb{L}_{s\in\mathbb{S}}^*$ }
 \label{alg_symbol_spread_2}
 }
\end{algorithm}
}
\vspace{0mm}
%
%
%
%
Algorithm~\ref{alg_symbol_spread_2} illustrates a simple and heuristic algorithm for finding $\mathbb{L}_{s\in\mathbb{S}}^*$.
\begin{example}
Consider again the ANS from Section~\ref{sec_ans} with $L=16$ states and three symbols $\mathbb{S}=\{s_1,s_2,s_3\}$ that occur with
probabilities $3/16$, $5/16$ and $8/16$, respectively.
We follow Algorithm~\ref{alg_symbol_spread_2} and get the following table:
\[\footnotesize
\arraycolsep=0.9pt\def\arraystretch{1.4}
\begin{array}{|c|c|c|c|c| c|c|c|c| c|c|c|c| c|c|c|c|} \hline
x\rightarrow & 16 & 17 & 18 & 19 & 20 & 21 & 22 & 23 & 24 & 25 & 26 & 27 & 28 & 29 & 30 & 31 \\ \hline
y\rightarrow &16.75&16.98&17.86&17.98 &19.95&19.98&21.98&22.56 &23.16&23.98&25.56&25.99 &27.91&27.99&28.77& 29.99\\ \hline
s\rightarrow &2&3&1&3 &2&3&3&1 &2&3&2&3 &1&3&2&3 \\ \hline
\end{array}
\]
After calculation of equilibrium probabilities, we get the average encoding length,
which is $\kappa=\frac{230755}{156048}\approx 1.4787$. \hfill $\Box$
\end{example}
\noindent
The following observations are relevant.
\begin{itemize}
\item The algorithm for tuning symbol spread is very inexpensive and can be easily applied for
	ANS with a large number of states (bigger than $2^{10}$). It gives a good compression rate.
\item Equation~\ref{eq_preferred} gives a rational number that needs to be rounded up or down.
	Besides, preferred states pointed by it are likely to collide.
	Algorithm~\ref{alg_symbol_spread_1} computes a symbols spread, which follows
	the preferred positions. 
\item Equation~\ref{eq_preferred} indicates that preferred states are likely to be uniformly distributed within $\mathbb{I}$.	
\item It seems that finding $\mathbb{L}_{s\in\mathbb{S}}^*$ such that 
	$d(\mathbb{L}_{s\in\mathbb{S}}^*, {\cal L}_{s\in\mathbb{S}})$ attains minimum
	does not guarantee the best compression rate. However, it results in a ANS instance with a ``good'' compression rate.
	This could be a starting point for searching ANS with a better compression rate.
\end{itemize}

\section{Optimisation of ANS}\label{sec_best_ans}


\subsection{Case Study}
Consider a toy instance of ANS with three symbols that occur with probabilities $\{3/16,5/16,8/16\}$
and 16 states (i.e. $R=4$).
We have implemented a software in PARI that exhaustive searches through all possible
instances of ANS symbol spreads ($\frac{16!}{8!\cdot 5!\cdot 3!}=720720$).
For each spread, we have calculated equilibrium probabilities for the corresponding Markov chain.
This allows us to compute average length of a symbol (in bits/symbol). The results are shown in Table~\ref{tab_example_ans}.
\begin{table}[h]
\caption{Distribution of average lengths of ANS encodings for the toy ANS. Note that $H(\mathbb{S})\approx 1.477$}
\vspace{-10mm}
\begin{center}
	\[
	\arraycolsep=1.9pt\def\arraystretch{1.4}
	\begin{array}{| c || c|c|c|c|c|c|c|} \hline
	\mbox{Total}\#	& Min & A & B & C &D &  Max & \#\mbox{Fail }\\  \hline
	720,720& 30,240 & 86,560 & 483,360 & 66,896 & 48,568 & 56 & 5,040\\ \hline
	\end{array}
	\]
\end{center}
{\footnotesize Legend: $A=\langle Min,1,48]$, $B= \langle 1.48,1.49]$,
$C= \langle1.49,1.5]$, $D= \langle1.5, Max\rangle$, $Min=\frac{3619}{2448}\approx 1.478$
and $Max=\frac{97}{64} \approx 1.515$.
}
\vspace{-2mm}
\label{tab_example_ans}
\end{table}%
Unfortunately, there is a small fraction of ANS instances, whose equilibrium probabilities are impossible to establish
as the corresponding system is linearly dependent (its rank is 15).
An example of symbol spread with the shortest average encoding length is:
{\small
\[
		\overline{s}(x)=
	\left\{
	\begin{array}{cl}
	  s_1 &  \mbox{ if } x\in\{16, 24, 25\} =\mathbb L_1 \\
	  s_2 & \mbox{ if } x \in \{17, 20, 21, 26, 27\}=\mathbb L_2\\
	  s_3 &  \mbox{ if } x\in\{18, 19, 22, 23, 28, 29, 30, 31\}=\mathbb L_3\\
	\end{array}
	\right. 
\]}
In contrast, the following symbol spread function has the longest average encoding length
{\small
\[
		\overline{s}(x)=
	\left\{
	\begin{array}{cl}
	  s_1 &  \mbox{ if } x\in\{24, 25, 26\} =\mathbb L_1 \\
	  s_2 & \mbox{ if } x \in \{27, 28, 29, 30, 31\}=\mathbb L_2\\
	  s_3 &  \mbox{ if } x\in\{16, 17, 18, 19, 20, 21, 22, 23\}=\mathbb L_3\\
	\end{array}
	\right. 
\]}
Clearly, a designer of ANS is likely to use a pseudorandom number generator to select
symbol spreads. There is a better than $1/2$ probability that a such
instance has the average encoding length somewhere in the interval $\langle 1.48, 1.49]$.
It is interesting to see how quickly such instance approaches the best cases.


\subsection{Optimal ANS for Fixed ANS Parameters}
An idea is to start from a random symbol spread.
Then, we continue swapping pairs of ANS states. After, each swap we calculate residual redundancy
of a new ANS instance. If the redundancy is smaller than for the old instance we keep the change.
Otherwise, we select a new pair of states for a swap.
Details are given in Algorithm~\ref{alg_optimal}.
{\small 
\begin{algorithm}[h] \label{alg_optimal}
 {\footnotesize
 \KwData{symbol probability distribution $\{p_s; s\in\mathbb{S}\}$ and parameter $R$ such that $L_s=p_s2^R$ for all $s$}
 \KwResult{ symbol spread $\overline{s}$ or encoding table $\mathbb{E}(s,x)$ of ANS with the smallest residual redundancy}
 \KwSteps{\\
$\bullet$ initialise symbol spread $\overline{s}$ using PRNG;\\
$\bullet$ determine the corresponding $\mathbb{E}(s,x)$ and calculate its redundancy $\Delta H$; \\
$\bullet$ assume a required minimum redundancy threshold $T$; \\
\While(){$\Delta H \geq T$}{
	\For(){$x=2^R,\ldots, 2^{R+1}-1$}{
	$y \leftarrow PRNG(\mathbb{I})$, i.e. random selection of state from $\mathbb{I}$;\\
	\If(){$\overline{s}(x) \neq \overline{s}(y)$}{
		$\bullet$ calculate equilibrium probabilities for ANS with $\;\;$ swapped states;\\
		$\bullet$ determine average encoding length and $\;\;$ $\Delta H_{temp}$;
		}
	\If(){$\Delta H_{temp} < \Delta H$}{
		$\bullet$ update $\overline{s}$ by swapping $x$ and $y$;\\
		}
	}
}
$\bullet$ return $\overline{s}$ and $\Delta H$;
}  
\caption{Search for Optimal ANS}
 }
\end{algorithm}
}
\vspace{0mm}
%
%
We have implemented the algorithm in the PARI/GP environment.
Our experiment is run for 16-state ANS and three symbols that occur with probabilities $\{3/16,5/16,8/16\}$.
As the algorithm is probabilistic, its behaviour varies depending on specific
coin tosses when choosing state swaps (or $\{24, 25, 26\}$,  $\{27, 28, 29, 30, 31\}$,  $\{16, 17, 18, 19, 20, 21, 22, 23\}$).
Starting symbol spread is the one with the longest average encoding length
as this is the worst case.
\begin{table}[h]
\caption{Number of Swaps when Searching for Optimal ANS Instances with 16 states and 3 symbols}
\vspace{-10mm}
\begin{center}
	\[
	\def\arraystretch{1.5}
	\begin{array}{| c | c|c|c|c|} \hline
	\mbox{Average of Swaps} & \mbox{Min of Swaps} & \mbox{Max of Swaps} & \mbox{Min$\#$ Good Swaps} &  \mbox{Max$\#$ Good Swaps} \\ \hline
	\approx 24 & 4 & 223 & 4 & 19 \\ \hline
	\end{array}
	\]
\end{center}
\vspace{-6mm}
\label{tab_opt_stat}
\end{table}%
The algorithm continues to swap state pairs until the average length equals to the minimum $\frac{3619}{2448}$
(the optimum compression).
We have run the algorithm $10^5$ times.
The results are presented in Table~\ref{tab_opt_stat}.
The main efficiency measure is the total number of swaps of ANS state pairs
that is required in order to achieve optimum compression.
Note that each state flip forces algorithm to redesign ANS, to compute its state equilibrium probabilities
and to evaluate average encoding length.
As the algorithm uses PRNG, the number of state swaps varies.
The algorithm works very well and successfully achieves the optimal compression every time.
In the table, we introduce ``good swaps''.
It means that any good swap produces ANS instance whose average encoding length gets smaller.
This also means that in the worst case, we need only 19 swaps to produce optimal ANS.
Optimality here is understood as the minimum residual redundancy.

Complexity of Algorithm~\ref{alg_optimal} is ${\cal O}(\theta L^3)$,
where $\theta$ is the number of iterations of the main loop of the algorithm.
The most expensive part is Gaussian elimination needed to find equilibrium probabilities.
It takes ${\cal O}(L^3)$ steps.
We assume that we need $\theta$ swaps to obtain a required redundancy with 
high probability. In other words, the algorithm becomes very expensive, when the number of
states $L$ is bigger than $2^{10}$.
This is a bad news. A good news is that a system of linear equations defining equilibrium probabilities
is sparse.
Consider a simple geometric symbol probability distribution $\{1/2,1/4,\ldots, 1/1024,1/1024\}$.
Assume that ANS has $L=2^{10}$ states. A simple calculation points that the matrix $M$ in the relation
$M\cdot X=B$ for Markov equilibrium has $\approx 99\%$ zeros.
It means that we can speed up search for optimal ANS in the following way:
\begin{itemize}
\item use a specialised algorithms for Gaussian elimination that target sparse systems.
	There are some mathematical packages (such as MatLab for example) that include
	such algorithms;
\item apply inversion of sparse matrices (such as this from~\cite{CK2021}, whose complexity
	is ${\cal O}(L^{2.21})$). First we solve $M\cdot X=B$ by computing $M^{-1}$.
	This allows us to find $X$. Now we swap two states that belong to different symbols.
	Now we need to solve $\tilde{M}\cdot \tilde{X}=B$, where $\tilde{M}$ is system that
	describes equilibrium after the swap. 
	We build $\tilde{M}^{-1}$.
	Now we translate $M\cdot X=B$ by first $I\cdot M\cdot X=B$ and then
	$\tilde{M}\tilde{M}^{-1}M\cdot X=B$, which produces $\tilde{M}\cdot \tilde{X}=B$,
	where $\tilde{X}=\tilde{M}^{-1}M\cdot X$.
	In other words, the solution $\tilde{X}$ can be obtained from $X$ by multiplying it by $\tilde{M}^{-1} M$.
\end{itemize}

Further efficiency improvements can be achieved by taking a closer look at swapping operation.
For the sake of argument, consider a swap of two states and their
matrices $M$ and $\tilde{M}$ that describe their Markov chain probabilities before
and after swap, respectively.
There are essentially two distinct cases, when the swap is
\begin{itemize}
\item simple, i.e. only two rows of $M$ are affected by the swap. Matrix entries outside the main diagonal
	are swapped but entries on the main diagonal need to be handle with care as they do not change
	if their values are $-1$. For instance,
	take ANS from Table~\ref{tab_example_1}. Let us swap states 25 and 26. Rows of $M$ before
	swap are
	\[\small
	\arraycolsep=1.4pt\def\arraystretch{1.4}
	\begin{array}{|l|cccc |cccc |cccc |cccc|}\hline
	&16 &17&18&19& 20&21&22&23& 24&25&26&27& 28&29&30&31 \\ \hline
	25\rightarrow	&0&0&0&0& \frac{3}{16}& \frac{3}{16}& \frac{3}{16}& \frac{3}{16}&0 &-1&0&0& 0&0&0& 0\\
	26\rightarrow	&\frac{5}{16}&\frac{5}{16}&0&0& 0& 0& 0& 0&0 & 0&-1&0& 0&0&0& 0 \\ \hline
	\end{array}
	\]
	Rows after the state swap are
	\[\small
	\arraycolsep=1.4pt\def\arraystretch{1.4}
	\begin{array}{|l|cccc |cccc |cccc |cccc|}\hline
	&16 &17&18&19& 20&21&22&23& 24&25&26&27& 28&29&30&31 \\ \hline
	25\rightarrow	&\frac{5}{16}&\frac{5}{16}&0&0& 0& 0& 0& 0&0 & -1&0&0& 0&0&0& 0 \\ 
	26\rightarrow	&0&0&0&0& \frac{3}{16}& \frac{3}{16}& \frac{3}{16}& \frac{3}{16}&0 &0&-1&0& 0&0&0& 0\\
	\hline
	\end{array}
	\]
	The matrix $\tilde{M}$ after swap can be obtained by $\tilde{M}=M\Delta$, where $\Delta$
	is a sparse matrix that translates $M$ into $\tilde{M}$.
	As argued above, equilibrium solution is $\tilde{X}=\Delta^{-1}X$, where $\Delta^{-1}=\tilde{M}^{-1} M$.
	We still need to calculate $\tilde{M}^{-1}$, however, it is possible to recycle part of computations
	done while computing $M^{-1}$. This is possible as $M$ and $\tilde{M}$ share the same entries
	(except the two rows that correspond to the state swap),
\item complex, i.e. more than two rows need to be modified. This occurs when a swap causes a cascade
	of swaps that are needed to restore the increasing order in their respective $\mathbb{L}_s$.
	For example, consider ANS from Table~\ref{tab_example_1} and swap states $22$ and $26$.
	\[\small
	\begin{array}{l}
	\mathbb{L}_1=\{18,22,25\} \\
	\mathbb{L}_2=\{19,20,23,26,28\} 
	\end{array}
	\stackrel{22\leftrightarrow26}{\xrightarrow{\hspace*{12mm}}}
	{\red
	\boxed{
	\begin{array}{l}
	\{18,26,25\} \\
	\{19,20,23,22,28\} \\
	\mbox{\it invalid ANS}
	\end{array}}
	}
	\stackrel{ \scriptsize
	\begin{array}{c}
	25\leftrightarrow26\\
	22\leftrightarrow23
	\end{array}
	}{\xrightarrow{\hspace*{12mm}}}
	\begin{array}{l}
	\{18,25,26\} \\
	\{19,20,22,23,28\}
	\end{array}
	\]
	To obtain a valid ANS instance, we need two extra swaps. 
	Calculation of $\tilde{M}^{-1}$ can still be supported by computations for $M^{-1}$
	but the matrices $M$ and $\tilde{M}$ differ on more than two rows.
\end{itemize}

\subsection{Optimalisation with Quantisation}
So far we have assumed that $L_s=p_s\cdot L$; $s\in\mathbb{S}$
or at least $L_s/L$ is a very close approximation of $p_s$.
However, this is not always true. 
In practice, there are two issues that need to be dealt with.
\begin{itemize}
	\item The first is the fact that $p_s\cdot L$ may have two ``good'' approximations
		when $\alpha_s < p_sL < (\alpha_s+1)$, where $\alpha_s\in\mathbb{N}^+$.
		So, we can choose $L_s\in\{\alpha_s,\alpha_s+1\}$.
		This occurs more frequently when the number $L$ is relatively small.
	\item The second issue happens when there is a tail of symbols whose probabilities
		are small enough so $p_s\cdot L < 1$. It means that that symbols have to be
		assigned to single states by symbol spread or $L_s=1$.
		This is equivalent to an artificial increase of symbol probabilities of the tail to $1/L$.
		Note that $\sum_{s\in\mathbb{S}} p_s=1$, which is equivalent to $\sum_{s\in\mathbb{S}} L_s=L$.
		Consequently, numbers of states in symbol spread of other symbols have to be reduced.
\end{itemize}
The research question in hand (also called quantisation) is defined as follows.
Given a symbol probability distribution ${\cal P}=\{p_s|s\in\mathbb{S}\}$.
How to design ANS so its compression is optimal (or close to it), where ANS is built for
the symbol probability distribution ${\cal Q}=\{q_s|s\in\mathbb{S}\}$,
where  ${\cal Q}$ approximates  ${\cal P}$, where $L_s=q_s\cdot L \in \mathbb{N}^+$,
$q_s\approx p_s$ and $n=|\mathbb{S}|$.

Let us consider the following algorithms.
\vspace{-1mm}
\begin{itemize}
	\item Exhaustive search through all possible quantised ANS, where an ANS instance
		is determined by a selection of $L_s \leftarrow \{\alpha_s,\alpha_s+1\}$, where $s\in\mathbb{S}$.
		For each selection, we run Algorithm~\ref{alg_optimal}. Finally, we choose the ANS instance
		with the best compression rate. Unfortunately, complexity of the algorithm is exponential or
		more precisely, ${\cal O}(2^n \theta L^3)$. A possible tail of $t$ symbols reduces complexity as
		all $t$ symbols are assigned a single state and the next $t$ low probability symbols are assigned
		$\alpha_s$ states. Higher $L_s=\alpha_s+1$ are ignored in order to compensate states
		that have been taken by the tail symbols.
		Note also that some of selections for $L_s$ must be rejected if $\sum_{s\in\mathbb{S}} L_s\neq L$.
	\item Best-fit quantised ANS. For all symbols with high probabilities, we choose $L_s=\alpha_s+1$.
		For symbols with low probabilities, we select $L_s=\alpha_s$. This choice is balanced by
		the number of symbols in the tail where $L_s=1$.
		The intuition behind the algorithm is
		the fact that selection of higher $L_s$ reduces average length of encodings and vice versa.
		This gives us a unique or close to it solution for $L_s$; $s\in\mathbb{S}$.
		Now we can run Algorithm~\ref{alg_optimal} to find the optimal (or close to it) ANS.
		The best-fit algorithm and  Algorithm~\ref{alg_optimal} share the same complexity.
\end{itemize}
\noindent
To recap our discussion, we make the following remarks:
\begin{itemize}
\item Finding ANS with optimal compression is of a prime concern to everybody who
	would like to either maximise communication bandwidth or remove as much
	redundancy as possible from binary frames.
	We model asymptotic behaviour of ANS by Markov chains.
	Calculated equilibrium probabilities allow us to precisely determine the average
	length of binary encodings and consequently ANS compression rate.
\item Search for optimal ANS can be done in two steps. (1) We run 
	Algorithm either~\ref{alg_symbol_spread_1} or ~\ref{alg_symbol_spread_2}
	that tunes symbol spread using approximation of state probabilities.
	The algorithm is very efficient and its complexity is ${\cal O}(L)$.
	Unfortunately, it does not guarantee that calculated ANS instance is optimal.
	(2) Next we execute Algorithm~\ref{alg_optimal}. Its initial symbol spread
	has been calculated in the previous step.
	The algorithm is probabilistic and attains an optimal (or close to it) ANS with a high probability.
\item Algorithm~\ref{alg_optimal} can be sped up by using a specialised sparse matrix inversion
	algorithm together with reusing computations from previous inversions.
	This allows us to find optimal or close to optimal ANS for the number $L$ of states 
	in the range $[2^{10},2^{12}]$. The range is the most used in practice.
\end{itemize}

\section{Cryptographic ANS}\label{sec_cryptographic}
Duda and Niemiec~\cite{DN2016} have proposed a randomised ANS, where symbols spread is chosen
at random.  To make it practical, the authors suggest to replace a truly random source by
a pseudorandom number generator (PRNG), which is controlled by a relatively short random seed/key.
Two communicating parties can agree on a common secret key $K$.
Both sender and receiver use it to select their symbol spread using PRNG controlled by $K$.
The sender can build an appropriate encoding table, while the receiver -- the matching decoding table.
Consequently, the parties can use compression as an encryption.
Although, such encryption does not provide a high degree of protection (especially against integrity
and statistics attacks -- see~\cite{CDMMNPP2021,CDMMNPP2022}), it could be used effectively in low-security applications.
A price to be paid is a complete lack of control over compression rate of ANS.
This weakness can be mitigated by applying our Algorithm~\ref{alg_optimal}.
\begin{table}[h]
\caption{Cryptographic ANS}
\begin{center}
\begin{tabular}{|lcl|} \hline
Sender &  & Receiver \\ \hline
Secret $K$ &  & Secret $K$\\
Public ${\cal P}=\{ p_s | s\in\mathbb{S}\}$ &
$\xrightarrow{\hspace*{4mm}}$
& Public ${\cal P}=\{ p_s| s \in\mathbb{S}\}$\\
$\bullet$ Run Algorithm~\ref{alg_symbol_spread_2} $\rightarrow$ $\{L_s | s\in\mathbb{S}\}$ && 
$\bullet$ Run Algorithm~\ref{alg_symbol_spread_2} $\rightarrow$ $\{L_s | s\in\mathbb{S}\}$\\
$\bullet$ Run Algorithm~\ref{alg_optimal} with PRNG$(K)$  
&&
$\bullet$ Run Algorithm~\ref{alg_optimal} with PRNG$(K)$  \\
 $\;\;\;$ $\rightarrow$ $\{L_s^{best} | s\in\mathbb{S}\}$ 
&&
 $\;\;\;$ $\rightarrow$ $\{L_s^{best} | s\in\mathbb{S}\}$ \\
$\bullet$ Design encoding table for $\{L_s^{best} | s\in\mathbb{S}\}$ &&
$\bullet$ Build decoding table for $\{L_s^{best} | s\in\mathbb{S}\}$\\
  \hline
\end{tabular}
\end{center}
\label{tab_encryption}
\vspace{-2mm}
\end{table}%
Note that the symbol spread  $\{L_s | p_s\in\mathbb{S}\}$ is public but
$\{L_s^{best} | p_s\in\mathbb{S}\}$ is secret as to reconstruct it,
an adversary needs to recover $K$ and execute Algorithm~\ref{alg_optimal}.
Cryptographic ANS is illustrated in Table~\ref{tab_encryption}.
Unlike the Duda-Niemiec ANS, it achieves optimal (or close to it) compression.
But the effective security level (the length of cryptographic key) is determined
by the number of ANS instances produced by Algorithm~\ref{alg_optimal}.
For instance, ANS from Table~\ref{tab_example_ans} guarantees 15-bit security
(or $\log_2{30240}\approx 15$).
For ANS with a large number of states, it is difficult to determine precise security
level.
But this may be acceptable for low-security applications.

\section{Experiments}\label{sec_experiments}
Algorithm~\ref{alg_optimal} has been implemented using the Go language and 
has been executed on a MacBook Pro with M1 chip. 
The algorithm has been slightly modified so it finds both the lower and upper bounds for $\Delta H$. 
The lower bound points ANS, which is close to optimal. 
In contrast, the upper bound shows ANS, whose residual redundancy $\Delta H$ is big (close to the worst cases).
Note that a random selection of spreads produces ANS instances, whose $\Delta H$s
fall somewhere between the bounds.
The following results have been obtained for $10^5$ iterations of the {\tt FOR} loop.
\begin{center}
\resizebox{\textwidth}{!}{
\begin{tabular}{|c|c|c|c|c|}
\hline
\# ANS States             & 128                       & 256                   & 512                       & 1024 \\\hline
$\Delta H_{min}$          & 1.5770736607301217e-05    & 4.186921602089555e-06 & 1.0470777729310043e-06    & 2.7868954788345945e-07 \\\hline
$\Delta H_{max}$          & 0.03468776044228061       & 0.029874932999659265  & 0.03482157561417498       & 0.030663638836101903 \\\hline
\# Spreads with $\Delta H_{min}$     & 22                        & 68                    & 138                       & 156 \\\hline
\# Spreads with $\Delta H_{max}$    & 149                       & 357                   & 701                       & 1044 \\\hline
Search Time for $\Delta H_{min}$ & 1m12s                     & 8m30s                 & 1h11m37s                  & 9h2m59s \\\hline
Search Time for $\Delta H_{max}$ & 1m23s                    & 8m13s                 & 1h9m58s                   & 8h47m5s \\\hline
\end{tabular}
}
\end{center}
We have increased the number of iterations of 
the {\tt FOR} loop to $n^2$ where $n$ is the number of states for $n=\{512,1024\}$.
The results are presented below.
\begin{center}
{
\begin{tabular}{|c|c|c|}
\hline
$n$                           & 512                       & 1024 \\\hline
\# Iterations                        & 262144                    & 1048576 \\\hline
\# Good Swaps              & 508                       & 792 \\\hline
$\Delta H_{min}$                          & 1.022260727401303e-06     & 2.5842341977444505e-07  \\\hline
$\Delta H_{min}$ after additional rounds  & 1.0222607271792583e-06    & 2.5842341977444505e-07  \\\hline
\# Additional Rounds                 & $~10^6$                   & $~10^5$  \\\hline
Better Spreads Found in Additional Rounds   & 8                         & 0  \\ \hline
Execution Time                              & 872m                      & 6115m \\\hline
\end{tabular}
}
\end{center}

We see that due to probabilistic nature of the Algorithm~\ref{alg_optimal} even after a large number of iterations, there is a non-zero chance for finding a spread with lower $\Delta H$. 

In practise, there are time restrictions imposed on the time needed for execution of 
Algorithm~\ref{alg_optimal}. The following results illustrate how much time is needed
between two consecutive good swaps that improve $\Delta H$. The number of iterations of
the {\tt FOR} loop is $10^{5}$.
\begin{center}
{
\begin{tabular}{|c|c|c|c|c|}
\hline
\# ANS States             & 128                       & 256                   & 512                       & 1024 \\\hline
Tunning Time & 54ms & 235ms & 300ms & 374ms \\ \hline
Optimisation Time & 1m12s & 8m30s & 1h11m37 & 9h2m59s \\ \hline
\# Good Swaps & 22 & 68 & 138 & 156  \\ \hline
Time between Two Good Swaps & 3s & 7s & 31s & 209s  \\ \hline
Average $\Delta H$ Gain per Good Swap & 3e-07 & 2.2e-08 & 2.4e-09 & 3.7e-010 \\ \hline
$\Delta H_{min}$ & 1.5e-05 & 4.1e-06 & 1.04e-06 & 2.78e-07 \\ \hline
\end{tabular}
}
\end{center}

Note that matrix inversion consititues the main computational overhead of Algorithm~\ref{alg_optimal}. 
The experiments presented above have applied a standard Gaussian elimination (GE) for matrix inversion, whose
complexity is ${\cal O}(L^3)$. 
Algorithm~\ref{alg_optimal} can be sped up by
(1) using a more efficient algorithm for sparse matrix inversion (SMI) and
(2) recycling computations from previous matrix inversions.
The table below gives complexity of Algorithm~\ref{alg_optimal} for
different matrix inversion algorithms and 
for classical and quantum computers.
\begin{center}
{
\begin{tabular}{|c|c|c|c|}
\hline
\multicolumn{3}{|c|}{Classical Computer} & Quantum Computer \\ \hline
GE &SMI~\cite{CK2021}& SMI~\cite{HMS2013} & GE~\cite{HHL2008} \\ \hline
${\cal O}(\theta L^3)$ &  ${\cal O}(\theta L^{2.21})$  &  ${\cal O}(\theta L\log{L})$ & ${\cal O}(\theta (\log{L})^3)$  \\ \hline
\end{tabular}
}
\end{center}
The experiments have confirmed that Algorithm~\ref{alg_optimal} works well and is practical
for $L < 128$.
 However, for a larger $L$, it gets slower and quickly becomes impractical.
 The good news is that there is a significant room for improvement by applying 
 fast sparse matrix inversion algorithms together with recycling computations from previous matrix inversions.
However, optimisation of Algorithm~\ref{alg_optimal} is beyond of the scope of this work.
Note that the algorithm becomes very fast when it uses quantum matrix inversion.

\section{Conclusions}\label{sec_conclusions}
The work addresses an important practical problem of compression quality of the ANS algorithm.
In the description of ANS, its symbol spread can be chosen at random. 
Each symbol spread has its own characteristic probability distribution of ANS states.
Knowing the distribution, it is possible to compute ANS compression rate or alternatively
its residual redundancy $\Delta H$.

We present two algorithms that allows a user to choose symbol spreads that minimise $\Delta H$.
Algorithm~\ref{alg_symbol_spread_1} determines an ANS instance (its symbol spread)
whose state probabilities follow the natural ANS state bias.
It is it fast even for $L> 2^{12}$ but unfortunately, it does not provide the minimal/optimal $\Delta H$.
Algorithm~\ref{alg_optimal} provides a solution. 
It is able to find minimal $\Delta H$ with a probability that depends on the number of random coin tosses
$\theta$. 

We have conducted an experiment for $L=16$ that shows the behaviour of average length of ANS encodings.
Further experiments have confirmed that matrix inversion creates a bottleneck in Algorithm~\ref{alg_optimal}
and makes it impractical for large $L$. 
An immediate remedy is an application of specialised algorithms for sparse matrix inversion together with
recycling computations from previous matrix inversions.
Development of a fast version of Algorithm~\ref{alg_optimal} is left as a part of our future research.

The main research challenge is, however, to construct ANS instances in such a way that their
minimum residual redundancy is guaranteed by design.
It means that we have to understand interplay between symbol spreads and their equilibrium probabilities.
This points to an interesting connection between ANS and random graphs~\cite{BB2001}.

{\footnotesize
\bibliographystyle{plain}
\bibliography{ANS}
}
{\small
\section*{Acknowledgments}
Josef Pieprzyk, Marcin Paw\l{}owski and
Pawe\l{} Morawiecki have been supported by 
Polish National Science Center (NCN) grant 2018/31/B/ST6/03003.
}
\end{document}